\documentclass[aps, prb, twocolumn, showpacs, amssymb, floatfix]{revtex4}
\usepackage{amsmath,graphics,epsfig, mathrsfs}
\usepackage{bm}
\usepackage{epigraph}

\newcommand{\mfk}{\mathfrak}
\newcommand{\msc}{\mathscr}

\newcommand{\tx}{\text}
\newcommand{\ti}{\textit}

\newcommand{\dn}{\downarrow}
\newcommand{\Dna}{\Downarrow}

\newcommand{\nn}{\nonumber}
\newcommand{\olr}{\overleftrightarrow}
\newcommand{\pat}{\partial}

\newcommand{\up}{\uparrow}
\newcommand{\Upa}{\Uparrow}

\newcommand{\alp}{\alpha}

\newcommand{\gm}{\gamma}

\newcommand{\lam}{\lambda}

\newcommand{\vps}{\varepsilon}

\newcommand{\sg}{\sigma}

\newcommand{\etal}{{\em et al.~}}
\newcommand{\ie}{{i.e.,~}}

\begin{document}
\title{Tunable magnetization relaxation in spin valves}
\author{Xuhui Wang}
\email{xuhui.wang@kaust.edu.sa}
\author{Aurelien Manchon}
\affiliation{Physical Science \& Engineering Division, KAUST, 
Thuwal 23955-6900, Kingdom of Saudi Arabia}
\date{\today}

\begin{abstract}
In spin values the damping parameters of the free layer are determined 
non-locally by the entire magnetic configuration.
In a dual spin valve structure that comprises a free layer embedded between two pinned layers, 
the spin pumping mechanism, in combination with the angular momentum conservation, 
renders the tensor-like damping parameters tunable by varying the interfacial and diffusive properties.
Simulations based on the Landau-Lifshitz-Gilbert phenomenology for a macrospin model 
are performed with the tensor-like damping 
and the relaxation time of the free layer magnetization is found to be largely dependent on while tunable  
through the magnetic configuration of the source-drain magnetization.
\end{abstract}
\maketitle

A thorough knowledge of magnetization relaxation holds the key to understand 
magnetization dynamics in response to applied fields \cite{llg}
and spin-transfer torques.
\cite{slonczewski-berger-1996, stt-exp}  
In the framework of Landau-Lifshitz-Gilbert 
(LLG) phenomenology, relaxation is well captured by 
the Gilbert damping parameter that is usually cited as a scalar quantity. 
As pointed out by Brown half a century ago,\cite{brown-pr-1963}  the Gilbert 
damping for a single domain magnetic particle is in general a tensor.

When a ferromagnetic thin film is deposited on a normal metal substrate, 
an enhanced damping has been observed ferromagnetic 
resonance experiments. \cite{mizukami-2001}
This observation is successfully explained by spin pumping: 
\cite{tserkovnyak-prl-2002,tserkovnyak-rmp-2005}
The slow precession of the magnetization
pumps spin current into the adjacent normal metal where 
the dissipation of spin current provides a non-local mechanism to the 
damping. The damping enhancement is found to be proportional to spin mixing 
conductance, a quantity playing key roles in the magneto-electronic circuit theory.
\cite{tserkovnyak-rmp-2005,brataas-pr-2005}

The pumped spin current $\bm{I}_{p}\propto \bm{M}\times\dot{\bm{M}}$
is always in the plane formed by the free layer magnetization direction $\bm{M}$ 
and the instantaneous axis about which the 
magnetization precesses. Therefore, in a single spin valve, 
when $\bm{M}$ is precessing around the source (drain) 
magnetization $\bm{m}$, the pumping current
is always in the plane of $\bm{m}$ and $\bm{M}$. \cite{tserkovnyak-prb-2003} 
Let us assume an azimuth
angle $\theta$ between $\bm{m}$ and $\bm{M}$. In such an \ti{in-plane} configuration,
the pumping current $\bm{I}_{p}$ has a component $I_{p}\sin\theta$ that is
parallel to $\bm{m}$.  The spin transfer torque acting on the source (drain) ferromagnet 
$\bm{m}$ is the component of spin current that is in the plane and
perpendicular to $\bm{m}$. To simplify the discussion, we consider it to be completely 
absorbed by $\bm{m}$. The longitudinal (to $\bm{m}$) component  experiences 
multiple reflection at the source (drain) contact, and cancels the damping torque by 
an amount proportional to $I_{p}\sin^{2}\theta$ but is still aligned along the direction of 
$\bm{M}\times\dot{\bm{M}}$. Therefore the total damping parameter has an angle 
$\theta$ dependence but still picks up a scalar (isotropic) form. This is the well-known 
dynamic stiffness explained by Tserkovnyak \etal \cite{tserkovnyak-prb-2003}
In the most general case, when the precessing axis of the free layer is  
mis-aligned with $\bm{m}$, there is always an \ti{out-of-plane} pumping torque 
perpendicular to the plane. In the paradigm of Slonczewski, 
this \ti{out-of-plane} component is not absorbed at the interface 
of the source (drain) ferromagnetic nodes, while the conservation of angular momentum
manifests it as a damping enhancement that shows the tensor 
form when installed in the LLG equation. 

Studies in lateral spin-flip transistors have 
suggested a tensor form for the enhanced damping parameters.\cite{wang-prb-2006}
In spin valves, works based on general scattering theory 
have discussed the damping in the framework 
of fluctuation-dissipation theorem \cite{foros-prb-2008-2009} and 
shown that the Gilbert damping tensor can be expressed 
using scattering matrices,\cite{brataas-prl-2008}
thus enabling first-principle investigation.\cite{starikov-prl-2010}  
But explicit analytical expressions of the damping tensor, its
dependence on the magnetic configuration as well as the material 
properties and particularly its impact on the magnetization relaxation 
are largely missing.  

In this paper, we investigate the Gilbert damping parameters of the free layer 
in the so-called dual spin valve (DSV). \cite{berger-jap-2003,fuchs-apl-2005,balaz-yan-prb}
We analyze the origin of the damping tensor and derive explicit analytical expressions of 
its non-local dependence on the magnetic configuration and materials properties. 
A generalization of our damping tensor to a continuous magnetic texture 
agrees well with the results in earlier works. 
Particularly, we show, in numeric simulations, that by tuning the magnetic configurations 
of the entire DSV, the relaxation time of the free layer can be increased or decreased.
      
\begin{figure}[tbh]
\centering
\includegraphics[trim = 0mm 0mm 0mm 0mm, clip, scale=0.45]{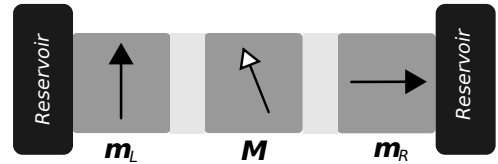}
\caption{A dual spin valve consists of a free layer (with magnetization direction 
$\bm{M}$) sandwiched by two fixed 
ferromagnetic layers (with magnetization directions $\bm{m}_{L}$ and $\bm{m}_{R}$)
through two normal metal spacers. The fixed layer are attached to reservoirs.}
\label{fig:dsv}
\end{figure}

To analyze the spin and charge currents in a DSV, we employ the 
magneto-electronic circuit theory and spin pumping,
\cite{tserkovnyak-rmp-2005,brataas-pr-2005}
in combination with diffusion equations.\cite{valet-fert-kovalev} 
Pillar-shaped metallic spin valves usually consist of normal-metal ($N$) spacers
much shorter than its spin-flip relaxation length, see for example
Ref.[\onlinecite{stt-exp,fuchs-apl-2005}]. 
To a good approximation, in the $N$ nodes, a spatially homogeneous spin accumulation is 
justified and the spin current ($\bm{I}_{i}$) conservation dictates  $\sum_{i}\bm{I}_{i}=0$ 
(where subscript $i$ indicates the source of spin current). 

A charge chemical potential ($\mu$) and a spin accumulation ($\bm{s}$) are
assigned to every $F$ or $N$ node. In a transition metal ferromagnet, 
a strong exchange field aligns the spin accumulation to the
magnetization direction. 
At every $F|N$ interface, the charge and spin currents on the $N$ side are 
determined by the contact conductance and 
the charge and spin distributions on both sides of the contact. For example,
at the contact between the left lead ferromagnet to the left normal metal $N_{1}$,
called $L|N_{1}$ thereafter, the currents are\cite{brataas-pr-2005} 
\begin{align}
I_{L} =& \frac{e}{2 h}G_{L}\left[(\mu_{1}-\mu_{L})
+P_{L}(\bm{s}_{1}-\bm{s}_{L})\cdot\bm{m}_{L}\right],\nn\\
\bm{I}_{L} =& -\frac{G_{L}}{8\pi}\left[2P_{L}(\mu_{1}-\mu_{L})\bm{m}_{L}
+(\bm{s}_{1}-\bm{s}_{L})\cdot\bm{m}_{L}\bm{m}_{L}\right.\nn\\
&\left.+\eta_{L}(\bm{s}_{1}-\bm{s}_{1}\cdot\bm{m}_{L}\bm{m}_{L})\right].
\end{align}
We have used the notation $G=g^{\up}+g^{\dn}$ is the sum of the spin-$\sg$ 
interface conductance $g^{\sg}$. The contact polarisation 
$P=(g^{\up}-g^{\dn})/(g^{\up}+g^{\dn})$. The ratio $\eta=2g^{\up\dn}/G$ is between 
the real part of the spin-mixing conductance $g^{\up\dn}$ and the total conductance $G$. 
The imaginary part of $g^{\up\dn}$ is usually much smaller than its real part, thus 
discarded.\cite{xia-prb-2002}
The spin-coherence length in a transition metal ferromagnet
is usually much shorter than the thickness of the thin film,\cite{stiles-prb-2002} 
which renders the mixing transmission negligible.\cite{tserkovnyak-rmp-2005}
The precession of the free layer magnetization $\bm{M}$ pumps a spin current 
$\bm{I}_{p}=(\hbar/4\pi)g_{F}^{\up\dn}\bm{M}\times\dot{\bm{M}}$ into the 
adjacent normal nodes $N_{1}$ and $N_{2}$, which is given by the mixing conductance
$g_{F}^{\up\dn}$ at the $F|N_{1(2)}$ interface (normal metals spacers are 
considered identical on both sides of the free layer).

A  back flow spin current at the $F|N_{1}$ interface reads 
\begin{align}
\bm{I}_{1} =& -\frac{G_{F}}{8\pi}\left[2P_{F}(\mu_{1}-\mu_{F})\bm{M}
+(\bm{s}_{1}-\bm{s}_{F})\cdot\bm{M}\bm{M}\right.\nn\\
&\left.+\eta_{F}(\bm{s}_{1}-\bm{s}_{1}\cdot\bm{M}\bm{M})\right]
\end{align}
on the $N_{1}$ side. Therefore, a weak spin-flip scattering in $N_{1}$ demands 
$\bm{I}_{L}+\bm{I}_{1}+\bm{I}_{p}=0$, which is dictated by angular 
momentum conservation. The same conservation law rules in $N_{2}$, where 
$\bm{I}_{R}+\bm{I}_{2}+\bm{I}_{p}=0$.

For the ferromagnetic ($F$) nodes made of transition metals, the spin diffusion 
is taken into account properly. 
\cite{tserkovnyak-prb-2003} In a strong ferromagnet, any transverse components 
decay quickly due to the large exchange field, thus the longitudinal spin accumulation 
$\bm{s}_{\nu}=s_{\nu}\bm{m}_{\nu}$  (with $\nu=L, R, F$) 
diffuses and decays exponentially at a length scale given by spin diffusion length 
($\lam_{sd}$) as $\nabla_{x}^{2}s_{\nu}=s_{\nu}/\lam_{sd}$.
The difference in spin-dependent conductivty of majority and minority carriers is 
taken into account by enforcing the continuity of longitudinal spin current 
$\bm{m}_{\nu}\cdot\bm{I}_{\nu}=-(D_{\nu}^{\up}\nabla_{x} s_{\nu}^{\up}
-D_{\nu}^{\dn}\nabla_{x} s_{\nu}^{\dn})$
at the every $F|N$ interface. We assume vanishing spin currents at the outer 
interfaces to reservoirs. 

The diffusion equations and current conservation determine, self-consistently, 
the spin accumulations and spin currents in both $N$ and $F$ nodes . 
We are mainly concerned with the exchange torque\cite{tserkovnyak-prb-2003} 
$\bm{T}=-\bm{M}\times(\bm{I}_{L}+\bm{I}_{R})\times\bm{M}$
acting on $\bm{M}$. 
A general analytical formula is attainable but lengthy. In the following, 
we focus on two scenarios that are mostly relevant to the state-of-the-art experiments
in spin valves and spin pumping: (1) The free layer has a strong spin flip (short $\lam_{sd}$)
and the thickness $d_{F}\geq \lam_{sd}$, for which the permalloy (Py) film is an 
ideal candidate;\cite{fuchs-apl-2005} (2) The free layer is a half metal, such as
Co$_{2}$MnSi studied in a recent experiment.\cite{chudo-jap-2011}  

\ti{Strong spin flip in free layer.}
We assume a strong spin flip scattering in the free layer \ie $d_{F}\geq \lam_{sd}$. 
We leave the diffusivity properties in the lead $F$ nodes arbitrary.
The total exchange torque is partitioned into two parts: An \ti{isotropic}
part that is parallel to the direction of the Gilbert damping 
$\bm{M}\times\dot{\bm{M}}$ and an \ti{anisotropic} part that is 
perpendicular to the plane spanned by $\bm{m}_{L(R)}$ and $\bm{M}$
(or the projection of $\bm{M}\times\dot{\bm{M}}$ to the direction 
$\bm{m}_{L(R)}\times\bm{M}$),  \ie
\begin{align}
\bm{T}=& \hbar\frac{g_{F}^{\up\dn}}{4\pi}
(\mfk{D}_{is}^{L}+\mfk{D}_{is}^{R})
\left(\bm{M}\times\dot{\bm{M}}\right)\nn\\
&+\hbar\frac{g_{F}^{\up\dn}}{4\pi}
\bm{M}\times\left[(\mfk{D}_{an}^{L}\hat{A}_{L,an}+\mfk{D}_{an}^{R}\hat{A}_{R,an})
\dot{\bm{M}}\right],
\label{eq:exchange-torque-strong-sf}
\end{align}
where the material-dependent parameters $\mfk{D}_{is}^{L(R)}$ and 
$\mfk{D}_{an}^{L(R)}$ are detailed in the Appendix \ref{sec:mater}.

Most interest is in the anisotropic damping described by  a symmetric tensor 
with elements
\begin{align}
\hat{A}_{an}^{ij}=-m_{i}m_{j}
\label{eq:aniso-damping}
\end{align}
where $i,j=x,y,z$ (we have omitted the lead index $L$ or $R$). 
The elements of $\hat{A}_{an}$ are given in Cartesian coordinates 
of the source-drain magnetization direction. The anisotropic damping appears as
$\bm{M}\times\hat{A}_{an}\dot{\bm{M}}$ that is always perpendicular to 
the free layer magnetization direction, thus keeping the length 
of $\bm{M}$ constant.\cite{foros-prb-2008-2009}
It is not difficult to show that when $\bm{M}$ is precessing around $\bm{m}$,
the anisotropic part vanishes due to $\hat{A}_{an}\dot{\bm{M}}=0$.

We generalize Eq.(\ref{eq:aniso-damping}) to a continuous magnetic texture.
Consider here only one-dimensional spatial dependence and the extension
to higher dimensions is straightforward. The Cartesian component of vector
$\bm{U}\equiv\bm{M}\times\hat{A}_{an}\dot{\bm{M}}$ is
$U_{i}=-\vps_{ijk}M_{j}m_{k}m_{l}\dot{M}_{l}$ (where $\vps_{ijk}$ is the 
Levi-Civita tensor and repeated indices are summed). We assume the fixed layer 
and the free layer differ in space by a lattice constant $a_{0}$, 
which allows $m_{k}\approx M_{k}(x+a_{0})$.
A Taylor expansion in space leads to 
$\bm{U}= -a_{0}^{2}\bm{M}\times(\hat{\mfk{D}}\dot{\bm{M}})$,
where the matrix elements $\hat{\mfk{D}}_{kl}=(\pat_{x}\bm{M})_{k}(\pat_{x}\bm{M})_{l}$
and we have assumed that the magnetization direction is always perpendicular 
to $\pat_{x}\bm{M}$. In this case, three vectors $\pat_{x}\bm{M}$, 
$\bm{M}\times\pat_{x}\bm{M}$ and $\bm{M}$ are perpendicular to each
other. A rotation around $\bm{M}$ by $\pi/2$  leaves $\bm{M}$ and $\dot{\bm{M}}$
unchanged while interchanging $\pat_{x}\bm{M}$ with
$\bm{M}\times\pat_{x}\bm{M}$, we have
\begin{align}
\hat{\mfk{D}}_{kl}=(\bm{M}\times\pat_{x}\bm{M})_{k}(\bm{M}\times\pat_{x}\bm{M})_{l},
\end{align}
which agrees with the so-called differential damping tensor Eq.(11) in 
Ref.[\onlinecite{zhang-prl-2010}].

Eq.(\ref{eq:exchange-torque-strong-sf}) suggests that the total exchange torque on the
free layer is a linear combination of two independent exchange torques arsing from
coupling  to the left and the right $F$ nodes. This form arises due to a
strong spin-flip scattering in the free layer that suppresses the exchange 
between two spin accumulations $\bm{s}_{1}$ and $\bm{s}_{2}$ in the $N$ nodes.  
In the pursuit of a concise notation for the 
Gilbert form, the exchange torque can be expressed as 
$\bm{T} = \bm{M}\times\olr{\alp}\dot{\bm{M}}$ with a total damping tensor 
given by
\begin{align}
\olr{\alp}=\hbar\frac{g_{F}^{\up\dn}}{4\pi}\left(\mfk{D}_{is}^{L}+\mfk{D}_{is}^{R}
+\mfk{D}_{an}^{L}\hat{A}_{L,an}+\mfk{D}_{an}^{R}\hat{A}_{R,an}\right).
\label{eq:total-damping}
\end{align}
The damping tensor $\olr{\alp}$ is determined by the entire
magnetic configuration of the DSV and particularly by the conductance
of $F|N$ contacts and the diffusive properties the $F$ nodes.  

\ti{Half metallic free layer}. This special while experimentally 
relevant \cite{chudo-jap-2011} case means
$P_{F}=1$. Half-metallicity in combination with the charge conservation 
enforces a longitudinal back flow that is determined solely by the bias current:
The spin accumulations in $N$ nodes do not contribute
to the spin accumulation inside the free layer, thus an independent contribution
due to left and right leads is foreseen.  
We summarize the material specific parameters in the Appendix \ref{sec:mater}.
When spin flip is weak in the source-drain ferromagnets,  $\xi_{L}\approx 0$
leads to $\mfk{D}_{is}\approx 0$. In this configuration, by taking
a (parallel or anti-parallel) source-drain magnetization direction as the precessing axis, the 
total damping enhancement vanishes, which reduces to the scenario of $\nu=1$ in 
Ref.[\onlinecite{tserkovnyak-prb-2003}].

\ti{Magnetization relaxation}.
To appreciate the impact of an anisotropic damping tensor on the magnetization relaxation, 
we perform a simulation, for the free layer magnetization, using Landau-Lifshitz-Gilbert 
(LLG) equation augmented by the tensor damping, \ie 
\begin{align}
\frac{d\bm{M}}{dt}=-\gm \bm{M}\times\bm{H}_{eff} 
& +\alp_{0}\bm{M}\times\frac{d\bm{M}}{dt}\nn\\
&+\frac{\gm}{\mu_{0} M_{s}V}\bm{M}\times\olr{\alp}\frac{d\bm{M}}{dt}.
\end{align}
$\alp_{0}$ is the (dimensionless) intrinsic Gilbert damping parameter. 
Symbol $\gm$ is the gyromagnetic ratio, $M_{s}$ is the saturation magnetization,
and $V$ is the volume of the free layer. $\mu_{0}$ stands for the vacuum permeability. 
The dynamics under the bias-driven spin transfer torque is not the topic in this paper, 
but can be included in a straightforward way.\cite{xiao-prb-2005} 
We give in the Appendix \ref{sec:stt}
the expressions of the bias-driven spin torques. 

We are mostly interested in the relaxation of the magnetization, instead of particular 
magnetization trajectories,  in the presence
of a tensor damping. The following simulation is performed for the scenario 
where the free layer has a strong spin flip, \ie Case (1). We employ the 
pillar structure from Ref.[\onlinecite{fuchs-apl-2005}] while considering the 
free layer (Py) to be 8nm thick (a thicker free layer favors a better thermal 
stability.\cite{fuchs-apl-2005})
The source-drain ferromagnets are cobalt (Co) and we expect the results are valid for 
a larger range of materials selections. The Py film is elliptic with three axes given by 
$2a=90~\tx{nm}$, $2b=35~\tx{nm}$, \cite{fuchs-apl-2005} and $c=8~\tx{nm}$. 
The demagnetizing factors $\msc{D}_{x,y,z}$ in the shape anisotropy energy 
$E_{\tx{dem}}=(1/2)\mu_{0}M_{s}^{2}V \sum_{i=x,y,z}\msc{D}_{i}M_{i}^{2}$
are $\msc{D}_{x}=0.50$, $\msc{D}_{y}=0.37$ and $\msc{D}_{z}=0.13$. 
An external field $\bm{H}_{a}$ leads to a Zeeman splitting
$E_{\tx{Zee}}=-V\mu_{0}M_{s}\bm{H}_{a}\cdot\bm{M}$. For Py films, we neglect the 
uniaxial anisotropy. The total free energy $E_{T}=E_{\tx{Zee}}+E_{\tx{dem}}$ gives rise 
to an effective field $\bm{H}_{eff}=-(1/VM_{s}\mu_{0})\pat E_{T}/\pat\bm{M}$.

The spin-dependent conductivities in the bulk of Co and the 
spin diffusion length $\lam_{Co}\approx 60~\tx{nm}$
are taken from the experimental data.\cite{bass-jmmm-1999}
For Py, we take $\lam_{Py}\approx 4~\tx{nm}$.\cite{fert-jmmm-1999}
To have direct connection with experiments, the above mentioned bare 
conductance has to be renormalized by the Sharvin conductance.\cite{bauer-prb-2003} 
For Py/Cu the mixing conductance, we take the value 
$g_{F}^{\up\dn}S^{-1}\approx 15~\tx{nm}^{-2}$,\cite{bauer-prb-2003}
which gives $\mfk{R}_{L(R)F}\approx 1.0$.

\begin{figure}[tbh]
\centering
\includegraphics[trim = 15mm 25mm 20mm 18mm, clip, scale=0.53]{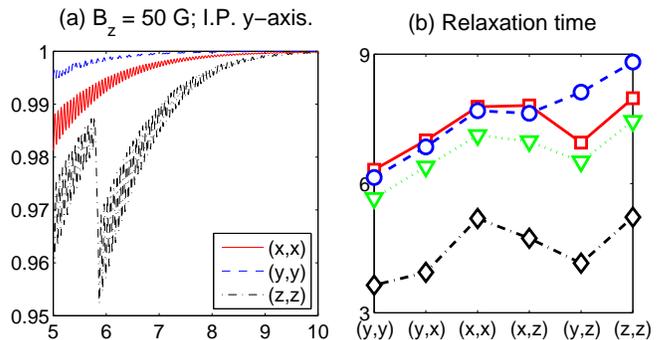}
\caption{(Color online) $M_{z}$ as a function of time (in ns) 
in presence of different source-drain magnetic configurations and applied fields.
(a) The external magnetic field $B_{z}=50~\tx{Gauss}$
is applied along $z$-axis. The blue (dashed), red (solid) and black (dotted dash) 
curves correspond to source-drain magnetization in configurations 
$(y, y)$, $(x,x)$, and $(z,z)$ respectively. 
(b) Magnetization relaxation times (in the unit of ns)versus source-drain magnetic configurations
at different applied field along $z$-axis.: $B_{z}=10~\tx{G}$ (red $\square$),  
$B_{z}=50~\tx{G}$ (blue $\bigcirc$), $B_{z}=200~\tx{G}$ (green $\bigtriangledown$), 
$B_{z}=800~\tx{G}$ (black $\Diamond$). Lines are a guide for the eyes. 
The initial position (I.P.) of the free layer is taken along $y$-axis.}
\label{fig:multiplot}
\end{figure}

The relaxation time  $\tau_{r}$ is extracted from the simulations by 
demanding at a specific moment $\tau_{r}$ the $|M_{z}-1.0|<10^{-3}$,
\ie reaches the easy axis. In the absence of bias, panel (a) of Fig.\ref{fig:multiplot} shows
the late stage of magnetization relaxation from an initial position ($y$-axis) in the presence 
of an tensor damping, under various source-drain (SD) magnetic configurations. 
The results are striking: Under the same field, switching the SD configurations 
increases or decreases $\tau_{r}$. 
In panel (b), the extracted relaxation times $\tau_{r}$ versus SD configurations 
under various fields are shown. At low field $B_{z}=10~\tx{G}$ (red $\square$), 
when switching from $(z,z)$ to $(y,y)$,
$\tau_{r}$ is improved from 8.0 ns to 6.3 ns, about 21\%. At a higher field 
$B_{z}=800~\tx{G}$ (black $\Diamond$), the improvement is larger from
5.2 at $(z,z)$ to 3.6 at $(y,y)$, nearly 31\%. To a large trend, the relaxation time
improvement is more significant at higher applied fields. 
 
In conclusion, combining conservation laws and 
magneto-electronic circuit theory, we have analyzed the Gilbert damping 
tensor of the free layer in a dual spin valve. Analytical results of the damping 
tensor as functions of the entire magnetic configuration and material properties 
are obtained. Numerical simulations on LLG equation augmented  
by the tensor damping reveal a tunable magnetization relaxation time by 
a strategic selection of source-drain magnetization configurations. Results presented
in this paper open a new venue to the design and control of magnetization 
dynamics in spintronic applications. 

X.Wang is indebted to G. E. W. Bauer, who has brought the
problem to his attention and offered invaluable comments.

\appendix 
\section{\label{sec:mater}Material dependent parameters}
In this paper, $\mfk{R}_{L(R)F}\equiv g_{L(R)}^{\up\dn}/g_{F}^{\up\dn}$ 
is the mixing conductance ratio and $\chi_{L(R)}\equiv \bm{m}_{L(R)}\cdot\bm{M}$. 
The diffusivity parameter 
$\xi_{L(R)}= \phi_{L(R)}(1-P_{L(R)}^{2})/\eta_{L(R)}$, where for the left $F$ node
\begin{align}
\phi_{L}=\frac{1}{1+\frac{(\sg_{L}^{\up}+\sg_{L}^{\dn})\lam_{L} e^{2}}
{4 h S \sg_{L}^{\up}\sg_{L}^{\dn}\tanh(d_{L}/\lam_{L})}
G_{L}(1-P_{L}^{2})}
\label{eq:phi-l}
\end{align}
where $h$ the Planck constant, $S$ the area of the thin film, $e$ the elementary charge,
$\lam_{L}$ the spin diffusion length, $d_{L}$ the thickness of the film, 
and $\sg^{\up(\dn)}$ the spin-dependent conductivity. $\phi_{R}$ is obtained 
by substituting all $L$ by $R$ in Eq.(\ref{eq:phi-l}). Parameter $\xi_{F}$ is given 
by  $\xi_{F}=(1-P_{F}^{2})\phi_{F}/\eta_{F}$ with
\begin{align}
\phi_{F}=\frac{1}{1+\frac{(\sg_{F}^{\up}+\sg_{F}^{\dn})\lam_{F} e^{2}}
{4 h S \sg_{F}^{\up}\sg_{F}^{\dn}}
G_{F}(1-P_{F}^{2})}.
\end{align}
The material dependent parameters as appearing in the damping tensor
Eq.(\ref{eq:total-damping}) are: (1) In the case of a strong spin flip in free layer,
\begin{align}
\mfk{D}_{is}^{L(R)} = & \frac{\mfk{R}_{L(R)F}}{\msc{L}_{L(R)F}}
\left[\xi_{L(R)}\mfk{R}_{L(R)F} +\xi_{L(R)}\xi_{F}(1-\chi_{L(R)}^{2})\right.\nn\\
&\left.+\xi_{F}(1-\chi_{L(R)}^{2})\chi_{L(R)}^{2}\right],\nn\\
\mfk{D}_{an}^{L(R)} = & \frac{\mfk{R}_{L(R)F}}{\msc{L}_{L(R)F}}
\frac{(\xi_{L(R)}-1)[\xi_{F}(1-\chi_{L(R)}^{2})+\mfk{R}_{L(R)F}]}{1+\mfk{R}_{L(R)F}},\nn\\
\msc{L}_{L(R)F}=& (1+\mfk{R}_{L(R)F}\chi_{L}^{2})\xi_{F}(1-\chi_{L(R)}^{2})\nn\\
&+\mfk{R}_{L(R)F} \left[(1-\chi_{L(R)}^{2})(1+\xi_{L(R)}\xi_{F})\right.\nn\\
&\left.+\xi_{L(R)}\mfk{R}_{L(R)F} 
+\xi_{F}\chi_{L(R)}^{2}\right];
\end{align}
(2)In the case of a half metallic free layer
\begin{align}
\mfk{D}_{is}^{L(R)}=& \frac{\mfk{R}_{L(R)F}\xi_{L(R)}}
{(1-\chi_{L(R)}^{2})+\xi_{L(R)}(\chi_{L}^{2}+\mfk{R}_{L(R)F})},\nn\\
\mfk{D}_{an}^{L(R)}=& \frac{\mfk{R}_{L(R)F}}{1+\mfk{R}_{L(R)F}}\nn\\
&\times\frac{\xi_{L(R)}-1}
{(1-\chi_{L(R)}^{2})+\xi_{L(R)}(\chi_{L}^{2}+\mfk{R}_{L(R)F})}.
\end{align}

\section{\label{sec:stt}Bias dependent spin torques}
The full analytical expression of bias dependent spin torques are rather lengthy. 
We give here the expressions, under a bias current $I$,  for symmetric 
SD ferromagnets  (\ie $\phi_{L}=\phi_{R}=\phi$ thus $\xi_{L}=\xi_{R}=\xi$)
with parallel or anti-parallel magnetization direction. 
(1) With a strong 
spin flip in the free layer, the parallel SD magnetization leads to vanishing 
bias-driven torque $\bm{T}_{\Upa\Upa}^{(b)}=0$; When the SD magnetizations 
are anti-parallelly (\ie $\bm{m}_{L}=-\bm{m}_{R}\equiv \bm{m}$), 
\begin{align}
\bm{T}_{\Upa\Dna}^{(b)}
=& \frac{I \hbar P\phi}{e (1+\mfk{R})\msc{L}}
\left[(\xi_{F}+\mfk{R}\xi_{F}\chi^{2}+\mfk{R})(1-\chi^{2})\right.\nn\\
&\left. +\mfk{R}(\mfk{R}+\xi_{F}(1-\chi^{2})+\chi^{2})\right]
\bm{m}_{F}\times(\bm{m}\times\bm{m}_{F}).
\end{align} 
(2) When the free layer is half metallic, for symmetric 
SD ferromagnets , $\bm{T}_{\Upa\Upa}^{(b)}=0$ and 
\begin{align}
\bm{T}_{\Upa\Dna}^{(b)}=\frac{I \hbar}{e}
\frac{\phi P}{(1-\xi)(1-\chi^{2})+\xi(\chi^{2}+\mfk{R})}
\bm{m}_{F}\times(\bm{m}\times\bm{m}_{F}).
\end{align}

\end{document}